\documentclass[12pt]{article}
\begin{document}
\title{The Thermodynamic Universe}
\author{B.G. Sidharth\\
Dipartimento di Mathematica e Informatica \\
Via delle Scienze 206,33100 Udine Italy\\ \&  Abdus Salaam International Center for Theoretical Physics \\Strada Costiera 11, Trieste 34104,Italy }
\date{}
\maketitle
\begin{flushleft}
{\large classification: 03.65.-w}\\
{\large keywords: Thermodynamic, Photon Mass, Gamma Rays.}
\end{flushleft}
\begin{abstract}
Using Planck scale oscillators in the background dark energy in a
model that parallels the theory of phonons, we deduce the Planck
mass, the elementary particle mass scale, the mass of the Universe
and a recently discovered residual energy in the cosmic background.
We also deduce the Beckenstein temperature formula for black holes.
Finally we show that the model explains the four minute time lag in
thearrival of gamma photons from a recently observed gamma flare by
the MAGIC telescope.
\end{abstract}
\section{Introduction}
The view we are trying to put forward is that the Universe is in
some sense immersed in a bath of "Dark Energy" and is
"Thermodynamic" in the sense that for example the temperature of
some material has a certain value. \\
To get a flavour, let us first derive the recently discovered
\cite{mersini} residual cosmic energy directly from the background
Dark Energy. We may reiterate that the "mysterious" background Dark
Energy is the same as the quantum Zero Point Fluctuations in the
background vacuum electromagnetic field.
Let us recall that the background Zero Point Field is a collection
of ground state oscillators \cite{mwt}. The probability amplitude is
$$\psi (x) = \left(\frac{m\omega}{\pi \hbar}\right)^{1/4} e^{-(m\omega/2\hbar)x^2}$$
for displacement by the distance $x$ from its position of classical
equilibrium. So the oscillator fluctuates over an interval
$$\Delta x \sim (\hbar/m\omega)^{1/2}$$
The background \index{electromagnetic}electromagnetic field is an
infinite collection of independent oscillators, with amplitudes
$X_1,X_2$ etc. The probability for the various oscillators to have
amplitudes $X_1, X_2$ and so on is the product of individual
oscillator amplitudes:
$$\psi (X_1,X_2,\cdots ) = exp [-(X^2_1 + X^2_2 + \cdots)]$$
wherein there would be a suitable normalization factor. This
expression gives the probability amplitude $\psi$ for a
configuration $B (x,y,z)$ of the magnetic field that is described by
the Fourier coefficients $X_1,X_2,\cdots$ or directly in terms of
the magnetic field configuration itself by,
$$\psi (B(x,y,z)) = P exp \left(-\int \int \frac{\bf{B}(x_1)\cdot \bf{B}(x_2)}{16\pi^3\hbar cr^2_{12}} d^3x_1 d^3x_2\right).$$
$P$ being a normalization factor. At this stage, we are thinking in
terms of energy without differenciation, that is, without
considering Electromagnetism or Gravitation etc as separate. Let us
consider a configuration where the magnetic field is everywhere zero
except in a region of dimension $l$, where it is of the order of
$\sim \Delta B$. The probability amplitude for this configuration
would be proportional to
$$\exp [-((\Delta B)^2 l^4/\hbar c)]$$
So the energy of \index{fluctuation}fluctuation in a region of
length $l$ is given by finally, the density \cite{mwt,uof}
$$B^2 \sim \frac{\hbar c}{l^4}$$
So the energy content in a region of volume $l^3$ is given by
\begin{equation}
\beta^2 \sim \hbar c/l\label{4e1}
\end{equation}
This energy is minimum when $l$ is maximum. Let us take $l$ to be
the radius of the Universe $\sim 10^{28}cms$. The minimum energy
residue of the background Dark Energy or Zero Point Field (ZPF) now
comes out to be $10^{-33}eV$, exactly the observed value. This
observed residual energy is a cosmic footprint of the ubiquitous
Dark Energy in the Universe, a puzzling footprint that, has recently been observed \cite{mersini}.\\
If on the other hand we take for $l$ in (\ref{4e1}) the smallest
possible length, the Planck length $l_P$, as
we will see in
the sequel, then we get the Planck mass $m_P \sim 10^{-5}gm$.\\
The minimum mass $\sim 10^{-33}eV$ or $10^{-65}gms$, has been shown by the author to
be the mass of the photon \cite{thermo}.Interestingly, this also is the minimum
thermodynamic mass in the Universe, as shown by Landsberg from a
totally different point of view, that of thermodynamics.  So (\ref{4e1}) gives two extreme
masses, the Planck mass and the photon mass. We will see how it is
possible to recover the intermediate elementary particle mass from
these considerations later.\\
As an alternative derivation, it is interesting to derive a model
based on the theory of phonons which are quanta of sound waves in a
macroscopic body \cite{huang}. In this
theory, as is well known the phonons have a maximum frequency
$\omega_m$ which is given by
\begin{equation}
\omega_m = c \left(\frac{6\pi^2}{v}\right)^{1/3}\label{4e2}
\end{equation}
In (\ref{4e2}) $c$ represents the velocity of sound in the specific
case of photons, while $v = V/N$, where $V$ denotes the volume and
$N$ the number of atoms. In this model we write
$$l \equiv \left(\frac{4}{3} \pi v\right)^{1/3}$$
$l$ being the inter particle distance. Thus (\ref{4e2}) now becomes
\begin{equation}
\omega_m = c/l\label{4e3}
\end{equation}
Let us now look upon it as a general set of Harmonic
oscillators as above. Then we can see that (\ref{4e3}) and
(\ref{4e1}) are identical.
So we again recover with suitable limits the extremes of the Planck
mass and the photon mass. (Other intermediate elementary particle
masses follow if we take $l$ as a typical Compton wavelength).\\
We now examine separately, the Planck scale.\\
Max Planck had noticed that, what we call the Planck scale today,
\begin{equation}
l_P = \left(\frac{\hbar G}{c^3}\right)^{\frac{1}{2}} \sim
10^{-33}cm\label{4ea1}
\end{equation}
is made up of the fundamental constants of nature and so, he
suspected it played the role of a fundamental length. Indeed, modern
Quantum Gravity approaches have invoked (\ref{4ea1}) in their quest
for a reconciliation of gravitation with other fundamental
interactions. In the process, the time honoured prescription of a
differentiable spacetime has to be abandoned.\\
There is also another scale made up of fundamental constants of
nature, viz., the well known Compton scale (or classical electron
radius),
\begin{equation}
l = e^2/m_ec^2 \sim 10^{-12}cm\label{4ea2}
\end{equation}
where $e$ is the electron charge and $m_e$ the electron mass.
Indeed if (\ref{4ea2}) is substituted for
$l$ in (\ref{4e1}), we get the elementary particle mass scale.\\
The scale (\ref{4ea2}) has also played an important role in modern
physics, though it is not considered as fundamental as the Planck
scale. Nevertheless, the Compton scale (\ref{4ea2}) is close to
reality in the sense of experiment, unlike (\ref{4ea1}), which is
well beyond foreseeable direct experimental contact. Moreover
another interesting feature of the Compton scale is that,
it brings out the Quantum Mechanical spin, unlike the Planck scale.\\
A very important question this throws up is that of a physical
rationale for a route from (\ref{4ea1}) to (\ref{4ea2}). Is there
such a mechanism?
\section{The Planck and Compton Scales}
 Our starting point \cite{bgsfpl172004,uof} is the
model for the underpinning at the Planck scale for the Universe.
This is a collection of $N$ Planck scale
oscillators where we will specify $N$ shortly.\\
 Let us consider an array of $N$
particles, spaced a distance $\Delta x$ apart, which behave like
oscillators that are connected by springs. We then have
\cite{bgsfpl152002,vandam,uof}
\begin{equation}
r  = \sqrt{N \Delta x^2}\label{4De1d}
\end{equation}
\begin{equation}
ka^2 \equiv k \Delta x^2 = \frac{1}{2}  k_B T\label{4De2d}
\end{equation}
where $k_B$ is the Boltzmann constant, $T$ the temperature, $r$ the
extent  and $k$ is the spring constant given by
\begin{equation}
\omega_0^2 = \frac{k}{m}\label{4De3d}
\end{equation}
\begin{equation}
\omega = \left(\frac{k}{m}a^2\right)^{\frac{1}{2}} \frac{1}{r} =
\omega_0 \frac{a}{r}\label{4De4d}
\end{equation}
We now identify the particles with \index{Planck}Planck
\index{mass}masses and set $\Delta x \equiv a = l_P$, the
\index{Planck}Planck length. It may be immediately observed that use
of (\ref{4De3d}) and (\ref{4De2d}) gives $k_B T \sim m_P c^2$, which
ofcourse agrees with the temperature of a \index{black hole}black
hole of \index{Planck}Planck \index{mass}mass. Indeed, Rosen
\cite{rosen} had shown that a \index{Planck}Planck \index{mass}mass
particle at the \index{Planck scale}Planck scale can be considered
to be a \index{Universe}Universe in itself with a Schwarzchild
radius equalling the Planck length. We also use the fact that a typical elementary particle like the \index{pion}pion can be
considered to be the result of $n \sim 10^{40}$ \index{Planck}Planck
\index{mass}masses.\cite{uof}\\
Using this in (\ref{4De1d}), we get $r \sim l$, the \index{pion}pion
\index{Compton wavelength}Compton wavelength as required. Whence the
pion mass is given by
$$m = m_P/\sqrt{n}$$
Further, in this latter case, using (\ref{4De1d}) and the fact that
$N = n \sim 10^{40}$, and (\ref{4De2d}),i.e. $k_BT = kl^2/N$ and
(\ref{4De3d}) and (\ref{4De4d}), we get for a \index{pion}pion,
remembering that $m^2_P/n = m^2,$
$$k_ B T = \frac{m^3 c^4 l^2}{\hbar^2} = mc^2,$$
which of course is the well known formula for the Hagedorn
temperature for \index{elementary particles}elementary particles
like \index{pion}pions .This
confirms that we can treat an elementary
particle as a series of some $10^{40}$ \index{Planck}Planck
\index{mass}mass
oscillators.\\
However it must be observed from (\ref{4De4d}) and (\ref{4De3d}),
that while the \index{Planck}Planck \index{mass}mass gives the
highest energy state, an elementary particle like the
\index{pion}pion is in the lowest energy state. This explains why we
encounter \index{elementary particles}elementary particles, rather
than \index{Planck}Planck \index{mass}mass particles in nature.
Infact as already noted \cite{bhtd}, a \index{Planck}Planck
\index{mass}mass particle decays via the \index{Bekenstein
radiation}Bekenstein radiation within a \index{Planck time}Planck
time $\sim 10^{-42}secs$.\\
Using the fact that the \index{Universe}Universe consists of $N \sim
10^{80}$ \index{elementary particles}elementary particles like the
\index{pion}pions, the question is, can we think of the
\index{Universe}Universe as a collection of $n N \, \mbox{or}\,
10^{120}$ Planck \index{mass}mass oscillators? Infact if we use equation (\ref{4De1d}) with
$$\bar N \sim 10^{120},$$
we can see that the extent is given by $r \sim 10^{28}cms$ which is
of the order of the diameter of the \index{Universe}Universe itself.
We shall shortly justify the value for $\bar{N}$. Next using
(\ref{4De4d}) we get
\begin{equation}
\hbar \omega_0^{(min)} \langle \frac{l_P}{10^{28}} \rangle^{-1}
\approx m_P c^2 \times 10^{60} \approx Mc^2\label{4De5d}
\end{equation}
which gives the correct \index{mass}mass $M$, of the
\index{Universe}Universe which in contrast to the earlier
\index{pion}pion case, is the highest energy state while the
\index{Planck}Planck oscillators individually are this time the
lowest in this description. In other words the
\index{Universe}Universe itself can be considered to be described in
terms of normal modes of
\index{Planck scale}Planck scale oscillators.\\
More generally, if an arbitrary mass $M$, as in (\ref{4De5d}), is
given in terms of $\bar{N}$ Planck oscillators, in the above model,
then we have from (\ref{4De5d}) and (\ref{4De1d}):
$$M = \sqrt{\bar{N}} m_P \, \mbox{and}\, R = \sqrt{\bar{N}} l_P,$$
where $R$ is the radius of the object. Using the fact that $l_P$ is
the Schwarzchild radius of the mass $m_P$, this gives immediately,
$$R = 2GM/c^2$$
a well known relation. In
other words, such an object, the Universe included as a special
case, shows up as a Black Hole.\\ In fact, we do not need to specify $N$. We have in this
case rewriting the relations (\ref{4De1d}) to (\ref{4De4d}),
$$R = \sqrt{N}l, Kl^2 = kT,$$
\begin{equation}
\omega^2_{max} = \frac{K}{m} = \frac{kT}{ml^2}\label{4ea3}
\end{equation}
In (\ref{4ea3}), $R$ is of the order of the diameter of the
Universe, $K$ is the analogue of spring constant, $T$ is the
effective temperature while $l$ is the analogue of the Planck
length, $m$ the analogue of the Planck mass and $\omega_{max}$ is
the frequency$--$the reason for the subscript $max$ will be seen
below. We do not yet give $l$ and $m$ their
usual values as given in (\ref{4ea1}) for example, but rather try to deduce these values.\\
 So using the Beckenstein temperature formula for
these primordial black holes \cite{ruffinizang}, that is
$$kT = \frac{\hbar c^3}{8\pi Gm}$$
in (\ref{4ea3}) we get,
\begin{equation}
Gm^2 \sim \hbar c\label{4e4}
\end{equation}
which is another form of (\ref{4ea1}). We can easily verify that
(\ref{4e4}) leads to the value $m \sim 10^{-5}gms$. In deducing
(\ref{4e4}) we have used the typical expressions for the frequency
as the inverse of the time - the Compton time in this case and
similarly the expression for the Compton length. However it must be
reiterated that no specific values
for $l$ or $m$ were considered in the deduction of (\ref{4e4}).\\
Finally we can see from (\ref{4ea3}) that, given the value of $l_P$
and using the value of the radius of the Universe, viz., $R \sim
10^{27}$, we can deduce that,
\begin{equation}
N \sim 10^{120}\label{4e7}
\end{equation}
The Compton scale (\ref{4ea2}) comes as a Quantum Mechanical effect,
within which we have zitterbewegung effects and a breakdown of
causal Physics as emphasized  elsewhere \cite{uof}.\\
We will now need a relation \cite{ijmpa}
\begin{equation}
G = \Theta/t\label{4ex}
\end{equation}
 This dependence also features in the Dirac
Cosmology and a few other
 cosmologies.\\
We now observe the following: It is known that for a Planck mass
$m_P \sim 10^{-5}gm$, all the energy is gravitational and in fact we
have, as in (\ref{4e4}),
$$Gm^2_P \sim e^2$$
To push these considerations further, we use the following relations for a
Schwarzchild black hole \cite{ruffinizang}:
$$dM = T d S, \, S = \frac{kc}{4\hbar G} A,$$
where $T$ is the Beckenstein temperature, $S$ the entropy and $A$ is
the area of the black hole. In our case, the mass $M = \sqrt{N}m_P$
and $A = Nl^2_P$, where $N$ is arbitrary for an arbitrary black
hole. Whence,
$$T = \frac{dM}{dS} = \frac{4\hbar G}{kl^2_Pc} \, \, \frac{dM}{dN}$$
If we use the fact that $l_P$ is the Schwarzchild radius for the
Planck mass $m_P$ and use the expression for $M$, the above reduces
to the Beckenstein temperature formula.\\
 Further, in this theory as is known \cite{ruffinizang},
\begin{equation}
\frac{dm}{dt} = - \frac{\beta}{m^2},\label{4e9}
\end{equation}
where $\beta$ is given by
$$\beta = \frac{\hbar c^4}{(30.8)^3 \pi G^2}$$
This leads back to the usual black hole life time given by, for any
mass $m$,
\begin{equation}
t = \frac{1}{3\beta} m^3 = 8.4 \times 10^{-24} m^3 secs\label{4ea7}
\end{equation}
Let us now factor in the time variation (\ref{4ex}) of $G$ into
(\ref{4e9}). Equation (\ref{4e9}) now becomes \cite{bhtd}
$$m^2 dm = -B \, \mu^{-2} t^2 dt, B \equiv \frac{\hbar c^4}{\lambda^3 \pi}, \,  \mu \equiv \frac{l c^2 \tau}{m}, \lambda^3 = (30.8)^3 \pi$$
Whence on integration we get
\begin{equation}
m = \frac{\hbar}{\lambda \pi^{1/3}} \left\{
\frac{1}{l^6}\right\}^{1/3} t \, = \frac{\hbar}{\lambda \pi^{1/3}}
\frac{1}{l^2} t\label{4e10}
\end{equation}
If we use the pion Compton time for $t$, in (\ref{4e10}), we get for
$m$, the pion mass. In other words, due to (\ref{4ex}), the
evanescent Planck mass decays into a stable elementary particle.\\
Before proceeding, we observe that we have deduced that a string of
$N$ Planck oscillators, $N$ arbitrary, form a Schwarzchild black
hole of mass $\sqrt{N}m_P = M$. Using the analogue of (\ref{4ea3})
and considerations before (\ref{4e9}), we can deduce that
$$\frac{dM}{dt} = m_P/t_P,$$
$$M = \left(m_P/t_P\right) \, \cdot \, t,$$
Whence $t$ being the "Beckenstein decay time" this is like the
equation (\ref{4e10}). For the Planck mass, $M = m_P$, the decay
time $t = t_P$. For the Universe, the above gives the life time $t$
as $\sim 10^{17}sec$, the age of the Universe! Interestingly, for
such black holes to have realistic life times $\geq 1 sec$, we
deduce that the mass $\geq 10^{38}gm$.\\
\section{Remarks}Recently the MAGIC gamma ray telescope in Spain detected a big gamma ray flare in the galaxy Mkn 501. It was observed that there was a four minute lag between the arrival times of gamma photons with energy greater than 10 Tev as compared to 100 Gev photons.This was sought to be explained by Ellis et al as a Quantum Gravity effect in which there is a dispersion in the velocities of the photons with respect to frequency \cite{arx}. They obtained
\[c' = c(1-\frac{E_0}{E'})\]
where $c'$ is the modified velocity of the photons and $E'$ is their estimate for the Planc energy, which turned out to be about one percent of the actual value.
Our above approach predicts a similar effect but this is due to the modified energy momentum dispersion relation deduced from theory \cite{uof},
\[ E^2 = E_0^2(1-\frac{E_0^2}{E''^2})\]
This gives
\[c'^2 = c^2(1-\frac{E_0^2}{E''^2})
\]
where $E''$ is the Planck energy.
On the other hand from the the equation above of Ellis et al. we get
\[c'^2 = c^2(1-\frac{E_0^2}{E''^2}.10^4)
\]
A comparison shows that our theory leads to the correct Planck energy unlike the Ellis et al value.

\end{document}